\documentstyle[12pt]{article}
\newlength{\extraspace}
\newlength{\extraspaces}
\newcommand{\be}{\begin{equation}
\addtolength{\abovedisplayskip}{\extraspaces}
\addtolength{\belowdisplayskip}{\extraspaces}
\addtolength{\abovedisplayshortskip}{\extraspace}
\addtolength{\belowdisplayshortskip}{\extraspace}}
\newcommand{\ee}{\end{equation}}

\newcommand{\ba}{\begin{eqnarray}
\addtolength{\abovedisplayskip}{\extraspaces}
\addtolength{\belowdisplayskip}{\extraspaces}
\addtolength{\abovedisplayshortskip}{\extraspace}
\addtolength{\belowdisplayshortskip}{\extraspace}}
\newcommand{\ea}{\end{eqnarray}}

\newcommand{\nonu}{\nonumber \\[.5mm]}
\newcommand{\A}{&\!\!\!}

\newcommand{\newsection}[1]{
\vspace{7mm} \pagebreak[3] \addtocounter{section}{1}
\setcounter{subsection}{0} \setcounter{footnote}{0}
\begin{center}
{\large {\bf \thesection. #1}}
\end{center}
\nopagebreak
\medskip
\nopagebreak \hspace{3mm}}

\begin{document}
\pagenumbering{arabic}

\begin{center}
{{\bf  Wormhole solution and Energy in Teleparallel Theory of
Gravity }}
\end{center}
\centerline{ Gamal G.L. Nashed}

\bigskip

\centerline{{\it Mathematics Department, Faculty of Science, Ain
Shams University, Cairo, Egypt }}

\bigskip
 \centerline{ e-mail:nashed@asunet.shams.edu.eg}

\hspace{2cm}
\\
\\
\\
\\
\\
\\
\\
\\

An exact  solution is obtained in the  tetrad theory of
gravitation. This solution is  characterized by two-parameters
$k_1,\ \ k_2$ of spherically symmetric static Lorentzian wormhole
which is obtained as a solution of the equation $\rho=\rho_t=0$
with $\rho=T_{i,j}u^iu^j$, $\rho_t=(T_{ij}-\displaystyle{1 \over
2}Tg_{ij}) u^iu^j$ where $u^iu_i=-1$. From this solution which
contains an arbitrary function we can generates the other two
solutions obtained before. The associated metric of this spacetime
 is a static Lorentzian wormhole and it includes the
Schwarzschild black hole, a family of naked singularity and a
disjoint family of Lorentzian wormholes. Calculate the energy
content of this tetrad field using the gravitational
energy-momentum given by M\o ller in teleparallel spacetime we
find that the resulting form depends on the arbitrary function and
does not depend on the two parameters $k_1$ and $k_2$ characterize
the wormhole. Using the regularized expression of the
gravitational energy-momentum we get the value of energy does not
depend on the arbitrary function.
\newpage
\begin{center}
\newsection{\bf Introduction}
\end{center}

At present, teleparallel theory seems to be popular again, and
there is a trend of analyzing the basic solutions of general
relativity with teleparallel theory and comparing the results.  It
is considered as an essential part of generalized non-Riemannian
theories such as the Poincar$\acute{e}$ gauge theory \cite{Yi1}
$\sim$ \cite{BN} or metric-affine gravity \cite{HMM} as well as a
possible physical relevant geometry by itself-teleparallel
description of gravity \cite{HS1,NH}.  Teleparallel approach is
used for positive-gravitational-energy proof \cite{Me1}. A
relation between spinor Lagrangian and teleparallel theory is
established \cite{TN}. In \cite{Lm5} it is shown that the
teleparallel equivalent of general relativity (TEGR) is not
consistent in presence of minimally coupled spinning matter. The
consistency of the coupling of the Dirac fields to the TEGR is
demonstrated
 \cite{ Me4}. However, it is shown that this demonstration is not
 correct \cite{OP,OP3}.

For a satisfactory description of the total energy of an isolated
system it is necessary that the energy-density of the
gravitational field is given in terms of first- and/or
second-order derivatives of the gravitational field variables. It
is well-known that there exists no covariant, nontrivial
expression constructed out of the metric tensor. However,
covariant expressions that contain a quadratic form of first-order
derivatives of the tetrad field are feasible. Thus it is
legitimate to conjecture that the difficulties regarding the
problem of defining the gravitational energy-momentum are related
to the geometrical description of the gravitational field rather
than are an intrinsic drawback of the theory \cite{Mj,MRTC}. M\o
ller has shown that the problem of energy-momentum complex has no
solution in the framework of gravitational field theories based on
Riemannian spacetime \cite{Mo}. In a series of papers,
\cite{Mo}$\sim$\cite{Mo1} he was able to obtain
 a general expression for a satisfactory energy-momentum complex in the teleparallel
  spacetime.

It was recognized by Flamm \cite{Fl} in (1916) that our universe
may not be simply connected, there may exist handles or tunnels
now called wormholes, in the spacetime topology linking widely
separated regions of our universe or even connected us with
different universes altogether. That such wormholes may be
traversable by humanoid travellers was first conjectured by Morris
and Thorne \cite{MT}, thereby suggesting that interstellar travel
and even time travel may some day be possible \cite{Vi,Kp}.

Morris and Thorne (MT) wormholes are static and spherically
symmetric and connect asymptotically flat spacetimes.  The metric
of this wormhole is given by \be
ds^2=-e^{2\Phi(r)}dt^2+\displaystyle{dr^2 \over 1-b(r)/r}+r^2(
d\theta^2+ \sin^2 \theta d\phi^2)\; , \ee where $\Phi(r)$ being
the redshift function and $b(r)$ is the shape function. The shape
function describes the spatial shape of the wormhole when viewed.
The metric (1) is spherically symmetric and static.  The geometric
significance of the radial coordinate $r$ is that the
circumference of a circle centered on the throat of the wormhole
is given by $2\pi r$. The coordinate $r$ is nonmonotonic in that
it decreases from $+ \infty$ to a minimum value $b_0$,
representing the location of the throat of the wormhole, and then
it increases from $b_0$ to $+ \infty$. This behavior of the radial
coordinate reflects the fact that the wormhole connects two
separate external universes. At the throat  $r=b=b_0$, there is a
coordinate singularity where the metric coefficient $g_{rr}$
becomes divergent but the radial proper distance \be l(r)=
\pm{\int_{b_0}}^r \displaystyle{ dr \over \sqrt{1-b(r)/r}},\ee
must be required to be finite everywhere \cite{Roa}. At the
throat, $l(r)=0$, while $l(r)<0$ on the left side of the throat
and $l(r)>0$ on the right side. For a wormhole to be traversable
it must have no horizon which implies that $g_{tt}$ must never
allowed to be vanish, i.e., $\Phi(r)$ must be finite everywhere.

Traversable Lorentzian have been in vogue ever since Morris, Thorn
and Yurtsever  \cite{MTY} came up with the exciting possibility of
constructing time machine models with these exotic objects. (MT)
paper demonstrated that the matter required to support such
spacetimes necessarily violates the null energy condition.
Semiclassical calculations based on techniques of quantum fields
in curved spacetime, as well as an old theorem of Epstein et. al.
\cite{EGY}, raised hopes about generation of such spacetimes
through quantum stresses.

There have been innumerable attempts at solving the "exotic matter
problem" in wormhole physics in the last few years \cite{Vi,VH}.
Alternative theories of gravity \cite{Hd} evolving wormhole
spacetimes \cite{Ks}$\sim$ \cite{HV1} with varying definitions of
the throat have been tried out as possible avenues of resolution.

It is the aim of the present work to derive a wormhole in M\o
ller's tetrad theory of gravitation. To do so we first begin with
a tetrad having spherical symmetry with three unknown functions of
the radial coordinate \cite{Ro}. Applying this tetrad to the field
equations of M\o ller's theory we obtain a set of non linear
partial differential equations. It is our aim to find a general
solution to these differential equations and discuss its physical
properties. In \S 2 a brief survey of M\o ller's tetrad theory of
gravitation is presented. The exact solution of the set of  non
linear partial differential equations is given in \S 3. In \S 4
the energy content of the tetrad field  is calculated and the form
of energy depends on the arbitrary function and does not depend on
the two parameters  $k_1$ and $k_2$ that characterize the wormhole
is obtained. In \S 5 the energy recalculated using the regularized
expression of the gravitational energy-momentum. Discussion and
conclusion of the obtained results are given in \S 6.
\newsection{M\o ller's  tetrad theory of gravitation}

In a spacetime with absolute parallelism the parallel vector
fields ${e_i}^\mu$ define the nonsymmetric affine connection \be
{\Gamma^\lambda}_{\mu \nu} \stackrel{\rm def.}{=} {e_i}^\lambda
{e^i}_{\ \mu, \ \nu}, \ee where ${e^i}_{\mu, \ \nu}=\partial_\nu
{e^i}_{\mu}$. The curvature tensor defined by
${\Gamma^\lambda}_{\mu \nu}$ is identically vanishing, however.

M\o ller's constructed a gravitational theory based on
 this spacetime. In this
theory the field variables are the 16 tetrad components
${e_i}^\mu$, from which the metric tensor is derived by \be g^{\mu
\nu} \stackrel{\rm def.}{=} \eta^{i j} {e_i}^\mu {e_j}^{\nu}, \ee
where $\eta^{i j}$ is the Minkowski metric $\eta_{i j}=\textrm
{diag}(+1\; ,-1\; ,-1\; ,-1).$

 We note that, associated with any tetrad field ${e_i}^\mu$ there
 is a metric field defined
 uniquely by (4), while a given metric $g^{\mu \nu}$ does not
 determine the tetrad field completely; for any local Lorentz
 transformation of the tetrads ${b_i}^\mu$ leads to a new set of
 tetrads which also satisfy (4).
  The Lagrangian ${\it L}$ is an invariant constructed from
$\gamma_{\mu \nu \rho}$ and $g^{\mu \nu}$, where $\gamma_{\mu \nu
\rho}$ is the contorsion tensor given by \be \gamma_{\mu \nu \rho}
\stackrel{\rm def.}{=} e_{i \ \mu }e_{i \nu; \ \rho}, \ee where
the semicolon denotes covariant differentiation with respect to
Christoffel symbols. The most general Lagrangian density invariant
under the parity operation is given by the form \cite{Mo1} \be
{\cal L} \stackrel{\rm def.}{=} \sqrt{-g} \left( \alpha_1 \Phi^\mu
\Phi_\mu+ \alpha_2 \gamma^{\mu \nu \rho} \gamma_{\mu \nu \rho}+
\alpha_3 \gamma^{\mu \nu \rho} \gamma_{\rho \nu \mu} \right), \ee
where \be g \stackrel{\rm def.}{=} {\rm det}(g_{\mu \nu}),
 \ee
 and
$\Phi_\mu$ is the basic vector field defined by \be \Phi_\mu
\stackrel{\rm def.}{=} {\gamma^\rho}_{\mu \rho}. \ee Here
$\alpha_1, \alpha_2,$ and $\alpha_3$ are constants determined by
M\o ller such that the theory coincides with general relativity in
the weak fields:

\be \alpha_1=-{1 \over \kappa}, \qquad \alpha_2={\lambda \over
\kappa}, \qquad \alpha_3={1 \over \kappa}(1-2\lambda), \ee where
$\kappa$ is the Einstein constant and  $\lambda$ is a free
dimensionless parameter\footnote{Throughout this paper we use the
relativistic units, $c=G=1$ and
 $\kappa=8\pi$.}. The same
choice of the parameters was also obtained by Hayashi and Nakano
\cite{HN}.

M\o ller applied the action principle to the Lagrangian density
(6) and obtained the field equation in the form \be G_{\mu \nu}
+H_{\mu \nu} = -{\kappa} T_{\mu \nu}, \qquad F_{\mu \nu}=0, \ee
where the Einstein tensor $G_{\mu \nu}$ is the Einstein tensor,
$H_{\mu \nu}$ and $F_{\mu \nu}$ are given by \be H_{\mu \nu}
\stackrel{\rm def.}{=} \lambda \left[ \gamma_{\rho \sigma \mu}
{\gamma^{\rho \sigma}}_\nu+\gamma_{\rho \sigma \mu}
{\gamma_\nu}^{\rho \sigma}+\gamma_{\rho \sigma \nu}
{\gamma_\mu}^{\rho \sigma}+g_{\mu \nu} \left( \gamma_{\rho \sigma
\tau} \gamma^{\tau \sigma \rho}-{1 \over 2} \gamma_{\rho \sigma
\tau} \gamma^{\rho \sigma \tau} \right) \right],
 \ee
and \be F_{\mu \nu} \stackrel{\rm def.}{=} \lambda \left[
\Phi_{\mu,\nu}-\Phi_{\nu,\mu} -\Phi_\rho \left({\gamma^\rho}_{\mu
\nu}-{\gamma^\rho}_{\nu \mu} \right)+ {{\gamma_{\mu
\nu}}^{\rho}}_{;\rho} \right], \ee and they are symmetric and skew
symmetric tensors, respectively.

M\o ller assumed that the energy-momentum tensor of matter fields
is symmetric. In the Hayashi-Nakano theory, however, the
energy-momentum tensor of spin-$1/2$ fundamental particles has
non-vanishing antisymmetric part arising from the effects due to
intrinsic spin, and the right-hand side of antisymmetric field
equation  (10) does not vanish when we take into account the
possible effects of intrinsic spin.

It can be shown \cite{HS1} that the tensors, $H_{\mu \nu}$ and
 $F_{\mu \nu}$, consist of only those terms which are linear or quadratic
in the axial-vector part of the torsion tensor, $a_\mu$, defined
by \be a_\mu \stackrel{\rm def.}{=} {1 \over 3} \epsilon_{\mu \nu
\rho \sigma} \gamma^{\nu \rho \sigma}, \qquad where \qquad
\epsilon_{\mu \nu \rho \sigma} \stackrel{\rm def.}{=} \sqrt{-g}
\delta_{\mu \nu \rho \sigma}, \ee where $\delta_{\mu \nu \rho
\sigma}$ being completely antisymmetric and normalized as
$\delta_{0123}=-1$. Therefore, both $H_{\mu \nu}$ and $F_{\mu
\nu}$ vanish if the $a_\mu$ is vanishing. In other words, when the
$a_\mu$ is found to vanish from the antisymmetric part of the
field equations (10), the symmetric part will coincides with the
Einstein field equation in teleparallel equivalent of general
relativity.
\newpage
\newsection{Spherically Symmetric Solutions}

Let us begin with the tetrad \cite{Ro}

\be \left({e_l}^\mu \right)= \left( \matrix{ A & Dr & 0 & 0
\vspace{3mm} \cr 0 & B \sin\theta \cos\phi & \displaystyle{B \over
r}\cos\theta \cos\phi
 & -\displaystyle{B \sin\phi \over r \sin\theta} \vspace{3mm} \cr
0 & B \sin\theta \sin\phi & \displaystyle{B \over r}\cos\theta
\sin\phi
 & \displaystyle{B \cos\phi \over r \sin\theta} \vspace{3mm} \cr
0 & B \cos\theta & -\displaystyle{B \over r}\sin\theta  & 0 \cr }
\right), \ee where {\it A}, {\it D}, {\it B}, are functions of the
radial coordinate $r$. The associated metric of the tetrad (14)
has the form \be ds^2=-\displaystyle{B^2-D^2r^2 \over
A^2B^2}dt^2-2\displaystyle{Dr \over AB^2}drdt+\displaystyle{1
\over B^2}dr^2+\displaystyle{r^2 \over B^2}(d\theta^2+ \sin^2
\theta d\phi^2). \ee As is clear from (15) that there is a cross
term which can be eliminated by performing the coordinate
transformation \be dT=dt+{ADr \over B^2-D^2r^2} dr, \ee using the
transformation (16) in the  tetrad (14) we obtain \be
\left({e_l}^\mu \right)= \left( \matrix{ \displaystyle {{\cal A}
\over 1-{\cal D}^2R^2} & (R{\cal D}-R^2{\cal D}{\cal B}') & 0 & 0
\vspace{3mm} \cr \displaystyle{{\cal A}{\cal D}R \sin\theta
\cos\phi \over 1-{\cal D}^2R^2}& (1-R{\cal B}')\sin\theta \cos\phi
& \displaystyle{\cos\theta \cos\phi \over R} &
-\displaystyle{\sin\phi \over R \sin\theta} \vspace{3mm} \cr
 \displaystyle{{\cal A}{\cal D}R \sin\theta \sin\phi \over
1-{\cal D}^2R^2} & (1-R{\cal B}') \sin\theta \sin\phi &
\displaystyle{\cos\theta \sin\phi \over R} &
\displaystyle{\cos\phi \over R \sin\theta} \vspace{3mm} \cr
\displaystyle{{\cal A}{\cal D}R \cos\theta  \over 1-{\cal D}^2R^2}
& (1-R{\cal B}') \cos\theta & -\displaystyle{\sin\theta  \over R}
& 0 \cr } \right), \ee where ${\cal A}$, ${\cal D}$ and ${\cal B}$
are now unknown functions of the new radial coordinates $R$ which
is defined by \be R=\displaystyle{r \over B}, \qquad \qquad {\cal
B}'=\displaystyle{d{\cal B}(R) \over dR}.\ee

Applying the tetrad fields (17) to the field equations (10)  we
obtain  the following non linear partial differential equations
 \ba
 \rho(R) \A= \A \displaystyle{1 \over \kappa}
 \Biggl[2\Biggl(R^3 {\cal D}^2{\cal B}'-R^2{\cal D}^2-R
{\cal B}'+1 \Biggr){\cal B}''+\Biggl(2R^3{\cal D}{\cal
D}'+5R^2{\cal D}^2-3\Biggr){\cal B}'^2\nonu
 \A \A
-2\Biggl(2R^2{\cal D}{\cal D}'
 +4R{\cal D}^2
 -\displaystyle{2 \over R} \Biggr){\cal B}'+\Biggl(2R {\cal D}'+3{\cal D}\Biggl){\cal D},\nonu
 \tau(R) \A= \A {1 \over \kappa R {\cal A}}\Biggl[
\Biggl(2R^3{\cal A}{\cal D}{\cal D}'-2R^3{\cal D}^2{\cal
A}'+3R^2{\cal A}{\cal D}^2+2R{\cal A}'-{\cal A}\Biggr)R{\cal B}'^2
-2\Biggl(2R^3{\cal A}{\cal D}{\cal D}'-2R^3{\cal D}^2{\cal A}'
\nonu
\A \A +3R^2{\cal D}^2{\cal A}+2R{\cal A}'-{\cal A}\Biggr){\cal
B}'+\Biggl(2R^2{\cal A}{\cal D} {\cal D}'-2R^2{\cal D}^2{\cal
A}'+3R{\cal A}{\cal D}^2+2{\cal A}'\Biggr)\Biggr], \nonu
 p(R) \A=\A  \kappa {T^3}_3= {1 \over \kappa R
{\cal A}^2}\Biggl[\Biggl({\cal A}R-{\cal A}{\cal D}^2R^5{\cal
B}'^2-{\cal A}{\cal D}^2R^3-2{\cal A}R^2{\cal B}'+2{\cal A}{\cal
D}^2R^4{\cal B}'+{\cal A}R^3{\cal B}'^2\Biggr){\cal A}''
 +\Biggl({\cal A}^2R\nonu
\A \A -{\cal A}^2R^4{\cal D}{\cal D}'+{\cal A}{\cal D}^2R^4{\cal
A}'-{\cal A}{\cal D}^2R^5{\cal A}'{\cal B}'+{\cal A}^2{\cal
D}R^5{\cal B}'{\cal D}'+{\cal A}R^3{\cal A}'{\cal B}'+2{\cal
A}^2{\cal D}^2R^4{\cal B}'-{\cal A}^2R^2{\cal B}'\nonu
\A \A -2{\cal A}^2{\cal D}^2R^3-{\cal A}R^2{\cal A}'\Biggr){\cal
B}''+\Biggl({\cal A}^2{\cal D}R^5{\cal B}'^2+{\cal A}^2{\cal
D}R^3-2{\cal A}^2{\cal D}R^4{\cal B}'\Biggr){\cal D}''+
\Biggl(2{\cal D}^2R^3-2R-4{\cal D}^2R^4{\cal B}'\nonu
\A \A +4R^2{\cal B}'\Biggr){\cal A}'^2+\Biggl({\cal A}^2R^5{\cal
D}'^2+2{\cal D}^2R^5{\cal A}'^2-{\cal A}^2R+2{\cal A}R^2{\cal
A}'+7{\cal A}^2{\cal D}R^4{\cal D}'-3{\cal A}{\cal D}R^5{\cal
A}'{\cal D}'-5{\cal A}{\cal D}^2R^4{\cal A}'\nonu
\A \A +5{\cal A}^2{\cal D}^2R^3-2R^3{\cal A}'^2\Biggr){\cal
B}'^2+\Biggl({\cal A}^2R^3-2{\cal A}^2R^4{\cal B}'\Biggr){\cal
D}'^2+\Biggl({\cal A}-4{\cal A}{\cal D}^2R^2\Biggr){\cal A}'
+\Biggr(9{\cal A}{\cal D}^2R^3{\cal A}'-3{\cal A}R{\cal A}'\nonu
\A \A-13{\cal A}^2{\cal D}R^3{\cal D}'+6{\cal A}{\cal D}R^4{\cal
A}'{\cal D}'+{\cal A}^2-8{\cal A}^2{\cal D}^2R^2\Biggr){\cal
B}'+\Biggl(6{\cal A}^2{\cal D}R^2-3{\cal A}{\cal D}R^3{\cal
A}'\Biggr){\cal D}'+3{\cal A}^2{\cal D}^2R \Biggr],
 \ea
where \[\rho(R)={T^0}_0, \qquad \qquad \tau(R)={T^1}_1, \qquad
\qquad p(R)={T^2}_2={T^3}_3,\] with $\rho(R)$ being the {\it
energy density}, $\tau(R)$ is the {\it radial pressure} and $p(R)$
is {\it the tangential pressure}. (Note that $\tau(R)$ as defined
above is simply the radial pressure $p_r$, and differs by a minus
sign from the conventions in \cite{MT,Vi}.)

Now let us try to solve the above differential equations (19).\\

 \underline {The General Solution}\\
It is our purpose to find a general solution to
 the differential equations  (19) when the stress-energy momentum tensor is not
 vanishing. From the first equation of (19) when $\rho(R)=0$, we can get
 the unknown function ${\cal D(R)}$ in terms of the unknown functions ${\cal B(R)}$
  to have the form \be
 {\cal D}(R)= \displaystyle{1 \over 1-R {\cal B}'}
\sqrt{\displaystyle{2m \over R^3}+ \displaystyle{{\cal B}' \over
R} \left(R {\cal B}' -2 \right)}, \ee  substitute (20) into (19)
we can obtain the  unknown function ${\cal A}(R)$ in terms of the
unknown function ${\cal} B(R)$ to have  form \be {
A(R)}=\displaystyle{1 \over \left(1-R {\cal
B}'\right)\left(k_2+\displaystyle{k_1 \over
\sqrt{1-\displaystyle{2m \over R}}}\right)}.\ee As is clear from
(20) and (21) that the  solution depends on the arbitrary function
${\cal B}$, i.e.,  we can generate the pervious solutions obtained
before by Nashed \cite{Nanc} by choosing the arbitrary function
${\cal B}$ to have the form \ba {\cal B}(R) \A =\A
\ln\left\{R\left(R-m+R\sqrt{1-\displaystyle{2m \over
R}}\right)\right\}-2\sqrt{1-\displaystyle{2m \over R}}, \qquad and
\qquad {\cal B}(R) = 1. \ea Using (20) and (21) in (10) we can get
the components of the energy-momentum tensor turn out to have the
form
 \be \rho(R)= 0,\qquad
 \tau(R)  = -\displaystyle{1 \over
 \kappa}\displaystyle{\left[2mk_1 \over
 R^3\left(k_1+k_2\sqrt{1-\displaystyle{2m \over R}}\right)
 \right]},\qquad
  p(R) = \displaystyle{1 \over
\kappa}\left[\displaystyle{mk_1 \over
R^3\left(k_1+k_2\sqrt{1-\displaystyle{2m \over R}}\right)}
\right]. \ee The weak \be \rho\geq 0, \qquad  \rho+\tau \geq 0,
\qquad  \rho+p \geq 0, \ee  and  null energy conditions  \be
\rho+\tau \geq 0, \qquad \rho+p \geq 0\ee  are both violated as is
clear from (23). The violation of the energy condition stems from
the violation of the inequality $\rho+\tau \geq 0$.

The complete line element of the above solution (20) and (21) is
\be ds^2= -\eta_1(R) dT^2 +{dR^2 \over \eta_2(R)} +R^2d\Omega^2,
\quad where \quad \eta_1(R)=\left(k_1+k_2\sqrt{1-\displaystyle{2m
\over R}} \right)^2, \quad  \eta_2(R)=\left({1-\displaystyle{2m
\over R}} \right), \ee with ${d\Omega^2=d\theta^2+\sin^2\theta
d\phi^2}$. If one replacing $k_2$ by $-k_2$ at the above solution
(20) and (21), the resulting form will also be a solution to the
non linear partial differential equations (19). All the above
solutions have a common property that their scalar Ricci tensor
vanishing identically, i.e., \[R(\{\})=0.\] The metric (26) makes
sense only for $R\geq 2m$ so to really  make the wormhole explicit
one needs two conditions patches\[ R_1 \in (2m ,\infty), \qquad
\qquad R_2 \in (2m ,\infty),\] which we then have to sew together
at $R=2m$. More discussion for such wormholes can be found in
\cite{DKV}. We are interested in the evaluation of energy since it
is the most important test for any gravitational energy
expression, local or quasi-local, since the geometrical setting
corresponds to an intricate configuration of the gravitational
field \cite{MRTC}.

\newsection{Energy content }

 The superpotential is given by \be {{\cal U}_\mu}^{\nu \lambda} ={(-g)^{1/2} \over
2 \kappa} {P_{\chi \rho \sigma}}^{\tau \nu \lambda}
\left[\phi^\rho g^{\sigma \chi} g_{\mu \tau}
 -\lambda g_{\tau \mu} \gamma^{\chi \rho \sigma}
-(1-2 \lambda) g_{\tau \mu} \gamma^{\sigma \rho \chi}\right], \ee
where ${P_{\chi \rho \sigma}}^{\tau \nu \lambda}$ is \be {P_{\chi
\rho \sigma}}^{\tau \nu \lambda} \stackrel{\rm def.}{=}
{{\delta}_\chi}^\tau {g_{\rho \sigma}}^{\nu \lambda}+
{{\delta}_\rho}^\tau {g_{\sigma \chi}}^{\nu \lambda}-
{{\delta}_\sigma}^\tau {g_{\chi \rho}}^{\nu \lambda} \ee with
${g_{\rho \sigma}}^{\nu \lambda}$ being a tensor defined by \be
{g_{\rho \sigma}}^{\nu \lambda} \stackrel{\rm def.}{=}
{\delta_\rho}^\nu {\delta_\sigma}^\lambda- {\delta_\sigma}^\nu
{\delta_\rho}^\lambda. \ee The energy is expressed by the surface
integral \cite{Mo2,MWHL,SNH} \be E=\lim_{r \rightarrow
\infty}\int_{r=constant} {{\cal U}_0}^{0 \alpha} n_\alpha dS, \ee
where $n_\alpha$ is the unit 3-vector normal to the surface
element ${\it dS}$.

Now we are in a position to calculate the energy associated with
solution (20) and (21) using the superpotential (27). As is
 clear from (30), the only components which contributes to the energy is ${{\cal U}_0}^{0
 \alpha}$. Thus substituting from solution (20) and (21) into
 (27) we obtain the following non-vanishing value
 \be
{{\cal U}_0}^{0 \alpha} ={2x^\alpha \over \kappa
r^3}\left(2m-R^2{\cal B}'+R^3{\cal B}'^2 \right).
 \ee
 Substituting from (31) into
(30) we get \be E(R)=2m-R^2{\cal B}'+R^3{\cal B}'^2. \ee We accept
the formula of the energy to depend on the physical quantities but
we do not accept the formula to depend on an arbitrary function
\cite{MRTC}.  Now we are going to follow a procedure similar to
that follow by  Brown-York formalism \cite{YB}.
\newsection{Regularized expression for the gravitational energy-momentum}

An important property of the tetrad fields that satisfy the
condition \be e_{i \mu}\cong \eta_{i \mu}+(1/2)h_{i \mu}(1/r),\ee
is that in the flat space-time limit
${e^i}_\mu(t,x,y,z)={\delta^i}_\mu$, and therefore the torsion
tensor  defined by
 \be {T^\lambda}_{\mu \nu}\stackrel{\rm
def.}{=}{e_a}^\lambda{T^a}_{\mu \nu}={\Gamma^\lambda}_{\mu
\nu}-{\Gamma^\lambda}_{\nu \mu},\ee  is vanishing, i.e.,
${T^\lambda}_{\mu \nu}=0$.  Hence for the flat space-time it is
normally to consider a set of tetrad fields such that
${T^\lambda}_{\mu \nu}=0$ {\it in any coordinate system}. However,
in general an arbitrary set of tetrad fields that yields the
metric tensor for the asymptotically flat space-time does not
satisfy the asymptotic condition given by (33). Moreover for such
tetrad fields the torsion ${T^\lambda}_{\mu \nu} \neq 0$ for the
flat space-time \cite{MRTC,MVR,MR}. It might be argued, therefore,
that the {\it expression for the gravitational energy-momentum
(30) is restricted to particular class of tetrad fields, namely,
to the class of frames such that ${T^\lambda}_{\mu \nu}=0$ if
${e^i}_\mu$ represents the flat space-time tetrad field}
\cite{MVR}. To explain this, let us calculate the flat space-time
tetrad field of (14) with (20) and (21) which is given by \be
\left({E_i}^\mu \right) =\left(\matrix {(1-R{\cal B}')
&\sqrt{R^2{\cal B}'^2-2R{\cal B}'} &0 &0 \vspace{3mm} \cr
\sqrt{R^2{\cal B}'^2-2R{\cal B}'} \sin\theta \cos\phi    &(1-R
{\cal B}') \sin\theta \cos\phi& \displaystyle{\cos\theta \cos\phi
\over R} &\displaystyle{- \sin\phi \over R\sin\theta} \vspace{3mm}
\cr \sqrt{R^2{\cal B}'^2-2R{\cal B}'} \sin\theta \sin\phi   &(1-R
{\cal B}') \sin\theta \sin\phi& \displaystyle{\cos\theta \sin\phi
\over R} &\displaystyle{ \cos\phi \over R\sin\theta} \vspace{3mm}
\cr \sqrt{R^2{\cal B}'^2-2R{\cal B}'} \cos\theta  &(1-R {\cal B}')
\cos\theta& \displaystyle{\sin\theta \over R} & 0
 \cr } \right). \ee Expression (35) yields
the following non-vanishing torsion components: \ba \A \A
T_{001}={\cal B}', \qquad T_{112}=-r\cos(\theta)\cos\phi {\cal
B}', \qquad T_{113}=\sin(\theta)\sin\phi {\cal B}', \qquad
T_{114}=-\displaystyle{\sin(\theta)\cos\phi \sqrt{{\cal
B}'}(1-{\cal B}') \over \sqrt{R^2{\cal B}'-2R}},\nonu
\A \A T_{124}=\cos\theta \cos\phi \sqrt{R^2{\cal B}'^2-2R{\cal
B}'}, \qquad T_{134}=-\sin(\theta)\sin\phi \sqrt{R^2{\cal
B}'^2-2R{\cal B}'}, \qquad T_{212}=-R\cos(\theta)\sin\phi {{\cal
B}'},\nonu
\A \A  T_{213}=-R\sin(\theta)\cos\phi {{\cal B}'}, \qquad
T_{214}=-\displaystyle{\sin\theta \sin\phi \sqrt{{\cal
B}'}(1-{\cal B}') \over \sqrt{R^2{\cal B}'-2R}},\quad
T_{224}=\cos(\theta)\sin\phi \sqrt{R^2{\cal B}'^2-2R{\cal
B}'},\nonu
\A \A  T_{234}=\sin(\theta)\cos\phi \sqrt{R^2{\cal B}'^2-2R{\cal
B}'}, \qquad T_{312}=R\sin(\theta){{\cal B}'}, \quad
T_{314}=-\displaystyle{\cos(\theta) \sqrt{{\cal B}'}(1-{\cal B}')
\over \sqrt{R^2{\cal B}'-2R}},\nonu
\A \A  T_{324}=-\sin(\theta)\sqrt{R^2{\cal B}'^2-2R{\cal B}'}.\ea

Maluf et al. \cite{MRTC,MVR,MR} discussed the above problem in the
framework of TEGR and constructed a regularized expression for the
gravitational energy-momentum in this frame. They checked this
expression for a tetrad field that suffer from the above problems
and obtain a very satisfactory results \cite{MVR}. In this section
we will follow the same procedure to derive a regularized
expression for the gravitational energy-momentum defined by Eq.
(30). It can be shown that one can defined  the gravitational
energy-momentum contained within an arbitrary volume $V$ of the
three-dimensional spacelike
 hypersurface  in the form \cite{Mo2,MWHL}
 \be P_\mu=\int_V d^3 x \partial_\alpha  {{\cal U}_\mu}^{0 \alpha},\ee  where
 ${{\cal U}_\mu}^{\nu \lambda}$ is given by Eq. (27).
 Expression (37) bears no relationship to the ADM
 (Arnowitt-Deser-Misner)
 energy-momentum \cite{MR}. $P_\mu$ transforms as a vector under the
 global SO(3,1) group.

 Our assumption is that the space-time be asymptotically flat. In
 this case the total gravitational energy-momentum is given by
 \be P_\mu=\oint_{S\rightarrow \infty} dS_\alpha \ {{\cal U}_\mu}^{0 \alpha}.\ee The
 field quantities are evaluated on a surface $S$ in the limit
 $r\rightarrow \infty$.

 In Eqs. (37) and (38) it is implicitly assumed that the reference space is determined
 by a set of tetrad fields ${e^i}_\mu$ for flat space-time such
 that the condition ${T^\lambda}_{\mu \nu}=0$ is satisfied. However, in
 general there exist flat space-time tetrad fields for which ${T^a}_{\mu \nu} \neq
 0$. In this case Eq. (37) may be generalized \cite{MVR,MR} by
 adding a suitable reference space subtraction term, exactly like
 in the Brown-York formalism \cite{YB}.

 We will denote ${T^a}_{\mu \nu}(E)=\partial_\mu {E^a}_\nu-\partial_\nu
 {E^a}_\mu$ and ${{\cal U}_\mu}^{0 \alpha} (E)$ as the expression of
 ${{\cal U}_\mu}^{0 \alpha} $
 constructed out of the flat tetrad ${E^i}_\mu$. {\it The
 regularized form of the gravitational energy-momentum $P_\mu$ is
 defined by}
 \be P_\mu=\int_{V} d^3x \partial_\alpha \left[ {{\cal U}_\mu}^{0 \alpha}(e)-
 {{\cal U}_\mu}^{0 \alpha} (E)\right].\ee
 This condition guarantees that the energy-momentum of
 the flat space-time always vanishes. The reference space-time is
 determined by tetrad fields ${E^i}_\mu$, obtained from
 ${e^i}_\mu$ by requiring the vanishing of the physical parameters
 like mass, angular momentum, etc. Assuming that the space-time is
 asymptotically flat then Eq. (39) can have the form

\be P_\mu=\oint_{S\rightarrow \infty} dS_\alpha \left[ {{\cal
U}_\mu}^{0 \alpha} (e)-{{\cal U}_\mu}^{0 \alpha} (E) \right],\ee
where the surface $S$ is established at spacelike infinity. Eq.
(40) transforms as a vector under the global SO(3,1) group
\cite{Mo1}. Now we are in a position to proof that the tetrad
field (14) with (20) and (21) yields a satisfactory value for the
total gravitational energy-momentum.

We will integrate Eq. (40) over a surface of constant radius
$x^1=r$ and require $r\rightarrow \infty$. Therefore, the index
$\alpha$ in (40) takes the value $\alpha=1$. We need to calculate
the quantity ${{\cal U}_0}^{0 1}$ and we find \be {{\cal U}_0}^{0
1} (e)\cong -\displaystyle{1 \over 4\pi}R\sin(\theta)({2m \over
R}-R{\cal B}'+R^2{\cal B}'^2),\ee and the expression of ${{\cal
U}_0}^{0 1} (E)$ is obtained by just making $m=0$ in Eq.(41). it
is given by \be {{\cal U}_0}^{0 1} (E)\cong-\displaystyle{1 \over
4\pi}R\sin(\theta)(R^2{\cal B}'^2-R{\cal B}').\ee Thus the
gravitational energy contained within a surface $S$ of constant
radius $r$ is given by \be P_0 \cong \int_{R\rightarrow \infty}
d\theta d\phi \displaystyle{1 \over 4\pi}\sin(\theta)\left\{-R({2m
\over R}-R{\cal B}'+R^2{\cal B}'^2)+(R^3{\cal B}'^2-R^2{\cal
B}')\right\}=2m,\ee this value of 2m is the value obtained by
several different approach \cite{MWHL,SNH}
\newsection{Discussion and conculusion}

In this paper we have applied the tetrad having spherical symmetry
with three unknown functions of the radial coordinate \cite{Ro} to
the field equations of M\o ller's tetrad theory of gravitation
\cite{Mo1}. From the resulting partial differential equation we
have  obtained an exact non vacuum solutions. This solutions is
characterized by an arbitrary function ${\cal B(R)}$ and from it
one can generates the other two solutions . The solutions in
general are characterize by three parameters $m$, $k_1$ and $k_2$.
If the two parameters $k_1=0$ and $k_2=1$ then one can obtains the
previous solutions \cite{Ngr}.  The energy-momentum tensor has the
property that $\rho=0$. The line element associated with these
solutions has the form (26).

To make the picture more clear we discuss the geometry of each
solution. The line element of this solution in the isotropic form
is  given by \[ ds^2= -\eta_1(R) dT^2 +{dR^2 \over \eta_2(R)}
+R^2d\Omega^2, \  where  \
\eta_1(R)=\left(k_1+k_2\sqrt{1-\displaystyle{2m \over R}}
\right)^2, \ \eta_2(R)=\left({1-\displaystyle{2m \over R}}
\right).\] If $g_{tt}=0$ one obtains a real naked singularity
region. Outside these regions naked singularity does not form and
one obtains a traverse wormhole. The throat of this wormhole
$g_{tt}(R=2m)$ gives the conditions that
$g_{tt}=-{k_1}^2\Rightarrow (k_1\neq0$ is required to ensure the
traversability).  The properties of this wormhole are discussed by
Dadhich et. al. \cite{DKV}.

We calculate the energy content of the solution (20) and (21)
using the energy-momentum complex given by \cite{Mo2,MWHL}. We
find that the energy does not depend on the two parameters $k_1$
and $k_2$ characterize the wormhole (32). On contrary it depends
on the arbitrary function ${\cal B(R)}$. This is in fact not
acceptable we accept the energy to depend on the physical
quantities like mass $m$ and the charge $q$ etc.

Maluf et al. \cite{MRTC,MVR,MR} have derived a simple expression
for the energy-momentum flux of the gravitational field. This
expression is obtained on the assumption that Eq.(37) represent
the energy-momentum of the gravitational field on a volume $V$ of
the three-dimensional spacelike hypersurface. They \cite{MVR,MR}
gave this definition for the gravitational energy-momentum in the
framework of TEGR, which require ${T^\lambda}_{\mu \nu}(E)=0$ for
the flat space-time. They extended this definition to the case
where the flat space-time tetrad fields ${E^a}_\mu$ yield
${T^\lambda}_{\mu \nu}(E) \neq 0$. They show that \cite{MR} in the
context of the regularized gravitational energy-momentum
definition it is not strictly necessary to stipulate asymptotic
boundary conditions for tetrad fields that describe asymptotically
flat space-times.

Using the definition of the torsion tensor given by Eq. (34) and
apply it to the tetrad field (35) we show that the flat space-time
associated with this tetrad field has a non-vanishing torsion
components Eq. (36). However, using the regularized expression of
the gravitational energy-momentum Eq. (40) and calculate all the
necessary components we finally get Eq. (43) which shows that the
total energy of the tetrad field (14) with (20) and (21) does not
depend on the arbitrary function.

As a punchline  we obtain a traversable  wormhole in  tetrad
theory of gravitation given by M\o ller \cite{Mo1} using a
spherical symmetric tetrad given by Robertson \cite{Ro} without
using the line element given by Eq. (1) \cite{MT}. Lemos et. al.
\cite{JLO} has studied Morris-Thorne wormholes with a cosmological
constant using the tetrad form of the line element (1) in the
diagonal form. Now one can do the same procedure with the non
diagonal tetrads given by the solutions (21) and (22).

\vspace{2cm} \centerline{\large {\bf Acknowledgment}} The author
 would like to thank  Professor J.G. Pereira  Universidade Estadual
 Paulista, Brazil.

\end{document}